\documentclass[10pt,letterpaper]{article}
\usepackage{opex3}
\usepackage{cite}

\begin{document}

\title{Weak measurements of a large spin angular splitting of light beam on
reflection at the Brewster angle}

\author{Xinxing Zhou, Hailu Luo$^{*}$, and Shuangchun Wen$^{\dagger}$}

\address{Key Laboratory for Micro-/Nano-Optoelectronic Devices of
Ministry of Education, College of Information Science and
Engineering, Hunan University, Changsha 410082, People's Republic of
China\nonumber\\ $^{\dagger}$scwen@vip.sina.com}

\email{$^{*}$hailuluo@hnu.edu.cn} 


\begin{abstract}
We reveal a large spin angular splitting of light beam on reflection
at the Brewster angle both theoretically and experimentally. A
simple weak measurements system manifesting itself for the built-in
post-selection technique is proposed to explore this angular
splitting. Remarkably, the directions of the spin accumulations can
be switched by adjusting the initial handedness of polarization.
\end{abstract}



\section{Introduction}
The spin Hall effect (SHE) of light manifests itself as a transverse
spin-dependent splitting, when a spatially confined light beam
passes from one material to another with different refractive
index~\cite{Onoda2004,Bliokh2006}. Recently, the transverse
splitting has been reported at an air-prism interface via weak
measurements~\cite{Hosten2008,Qin2009,Aiello2008}. On an
air-semiconductor interface, the transverse splitting has been
detected via ultrafast pump-probe techniques~\cite{Menard2009}. At
an air-metal interface, the transverse splitting has been reported
for the use of weak measurements and lock-in amplifying
methods~\cite{Zhou2012,Hermosa2011}. More recently, an in-plane
spin-dependent splitting has been observed when a linearly polarized
Gaussian beam impinges upon an air-prism interface~\cite{Qin2011}.
The spin-dependent splitting is generally believed as a result of an
effective spin-orbital coupling known as the influence of the
intrinsic spin (polarization) on the trajectory, which produces
transverse deflection of the spin. However, among these systems, the
spin-dependent splitting is tiny and reaches just a fraction of the
wavelength, limiting its future application.

In the present paper, we reveal a large spin angular splitting when
a slightly elliptical polarization beam incidents at the Brewster
angle. The reflected beam splits into two beams with opposite spin
polarizations propagating at different angles. As a result, we can
angularly sperate the beam with different polarizations. It should
be mentioned that the angular splitting is significantly different
from that in previous works where the splitting is limited to the
light intensity~\cite{Chan1985,Merano2009,Merano2010}. As an analogy
of SHE in semiconductor microcavity~\cite{Leyder2007}, the
directions of spin accumulations can be switched by adjusting the
initial handedness of polarization.

Importantly, this large spin angular splitting is explored with an
interesting simple weak measurements system which is similar to that
of Ref.~\cite{Gorodetski2012}. They propose that the coupling of a
tightly focused optical beam to surface plasmon polaritons offers a
natural weak measurements tool with a built-in post-selection. In
our work, the combination of the slightly elliptical polarization
incident beam with the reflected light beam at the Brewster angle
also provides a built-in post-selection weak measurements technique
for us. This simple weak measurements method shows significant
difference from the previous works in which an additional
post-selection is needed~\cite{Hosten2008,Qin2009,Aharonov1988}.
This spin angular splitting is also different from our previous
work~\cite{Luo2011b} where a linear polarization light beam
incidents near the Brewster angle and a large spatial shift is
observed. In addition, our work should be distinguished from the
angular Imbert-Fedorov shift~\cite{Hermosa2011a,Hermosa2012} whose
angular splitting is in the orthogonal plane of incidence.

\begin{figure}
\centerline{\includegraphics[width=7.5cm]{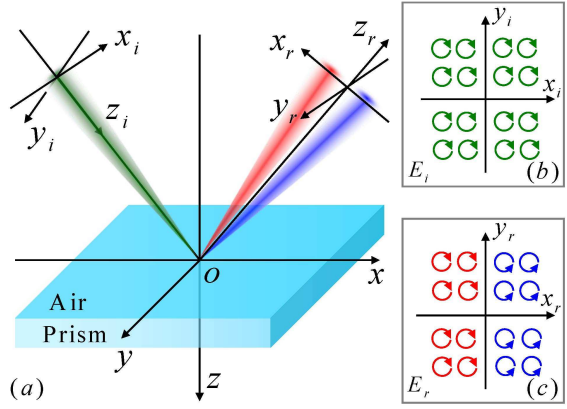}}
\caption{\label{Fig1} (a) Schematic illustrating the spin angular
splitting compared to the geometrical-optics prediction. (b) The
incident beam has an uniform polarization direction in the cross
section. (c) The polarization handedness experience different
rotations in reflection to satisfy transversality. }
\end{figure}

\section{Theoretical analysis}
The spin angular splitting is schematically shown in
Fig.~\ref{Fig1}(a). The $z$-axis of the laboratory Cartesian frame
($x,y,z$) is normal to the air-prism interface. We use the
coordinate frames ($x_i,y_i,z_i$) and ($x_r,y_r,z_r$) to denote
incidence and reflection, respectively. A left- or
right-elliptically polarized light beam incidents at the air-prism
interface. Here, we choose the long and short axis of the elliptical
polarization beam along to the $x_{i}$- and $y_{i}$-axis,
respectively. The elliptical polarization light beam can be
decomposed into two orthogonal polarization components $H$ and $V$.
It is noted that the mechanism of the elliptical polarization beam
reflection on the prism at the Brewster angel acts as a built-in
post-selection in which the $H$ component is mainly cut off and is
equal to the $V$ component. That is to say, this mechanism takes the
role of the second polarizer in the precious weak measurements
technique. After reflection, the $H$ and $V$ components overlap and
induce the large spin angular splitting. Additionally, the
reflection coefficient of $H$ polarization component changes its
sign across the Brewster angle, which means the induced total
circular polarization reverses its handedness [Fig.~\ref{Fig1}(b)
and~\ref{Fig1}(c)]. In other words, the two reflected beams are
"colored" by different circular polarization. Selection of circular
polarization in these beams is the post-selection procedure in the
weak-measurement technique.

We theoretically analyze the spin angular splitting with a general
beam propagation model. The reflected field $\tilde{\mathbf{E}}_r$
is related to the incident angular spectrum $\tilde{\mathbf{E}}_i$
by means of the relation~\cite{Bliokh2006,Luo2011a}
\begin{eqnarray}
\left[\begin{array}{cc}
\tilde{\mathbf{E}}_r^H\\
\tilde{\mathbf{E}}_r^V
\end{array}\right]
=\left[
\begin{array}{cc}
r_p&\frac{k_{ry} (r_p+r_s) \cot\theta_i}{k_0} \\
-\frac{k_{ry} (r_p+r_s)\cot\theta_i}{k_0} & r_s
\end{array}
\right]\left[\begin{array}{cc}
\tilde{\mathbf{E}}_i^H\\
\tilde{\mathbf{E}}_i^V
\end{array}\right]\label{matrixr}.
\end{eqnarray}
Here, $H$ and $V$ represent horizontal and vertical polarization
components, respectively. $\theta_i$ is the incident angle, $r_p$
and $r_s$ denote the Fresnel reflection coefficients for parallel
and perpendicular polarizations, respectively. $k_{0}$ is the wave
number in free space.

In the spin basis set, the incident angular spectrum for $H$ and $V$
polarizations can be written as:
$\tilde{\mathbf{E}}_i^H=(\tilde{\mathbf{E}}_{i+}+\tilde{\mathbf{E}}_{i-})/{\sqrt{2}}$
and
$\tilde{\mathbf{E}}_i^V=i(\tilde{\mathbf{E}}_{i-}-\tilde{\mathbf{E}}_{i+})/{\sqrt{2}}$.
Here,
$\tilde{\mathbf{E}}_{i+}=(\mathbf{e}_{ix}+i\mathbf{e}_{iy})\tilde{E}_{i}/\sqrt{2}$
and
$\tilde{\mathbf{E}}_{i-}=(\mathbf{e}_{ix}-i\mathbf{e}_{iy})\tilde{E}_{i}/\sqrt{2}$
denote the left- and right-circular polarized (spin) components,
respectively. We consider the incident beam with a Gaussian
distribution and its angular spectrum can be written as
\begin{equation}
\tilde{E}_{i}=\frac{w_0}{\sqrt{2\pi}}\exp\left[-\frac{w_0^2(k_{ix}^2+k_{iy}^2)}{4}\right]\label{asi},
\end{equation}
where $w_0$ is the beam waist. The reflected angular spectrum can be
obtained from Eq.~(\ref{matrixr}). In the spin basis,
$\tilde{\mathbf{E}}_r^H=(\tilde{\mathbf{E}}_{r+}+\tilde{\mathbf{E}}_{r-})/{\sqrt{2}}$,
$\tilde{\mathbf{E}}_r^V=i(\tilde{\mathbf{E}}_{r-}-\tilde{\mathbf{E}}_{r+})/{\sqrt{2}}$.
Here,
$\tilde{\mathbf{E}}_{r+}=(\mathbf{e}_{rx}+i\mathbf{e}_{ry})\tilde{E}_{r}/\sqrt{2}$
and
$\tilde{\mathbf{E}}_{r-}=(\mathbf{e}_{rx}-i\mathbf{e}_{ry})\tilde{E}_{r}/\sqrt{2}$,
where $\tilde{E}_{r}$ can be obtained from the boundary conditions:
$k_{ix}=-k_{rx}$ and $k_{iy}=k_{ry}$.

As for the elliptical polarization incident light beam, the Jones
vector can be written as $(\cos\Delta, e^{i\varphi}\sin\Delta)^{T}$.
Here $\Delta$ represents the azimuth angle (the angle between the
crystal axis of wave plate and the $x_{i}$-axis) and $\varphi$
denotes the phase difference between the two polarization components
$H$ and $V$. In the present study, we consider a elliptical
polarization beam with its long and short axis along to the $x_{i}$-
and $y_{i}$-axis. Therefore the Jones vector will be simplified to
$(\cos\Delta, +i\sin\Delta)^{T}$ or $(\cos\Delta, -i\sin\Delta)^{T}$
representing the left- or right-elliptical polarization in the case
of angle $\varphi=\pm \pi/2$. And we note here that the azimuth
angle $\Delta$ mentioned in the following is a tiny value allowing
for a slightly elliptical polarization and its long axis along to
the $x_{i}$-axis.
\begin{figure}[b]
\centerline{\includegraphics[width=7.5cm]{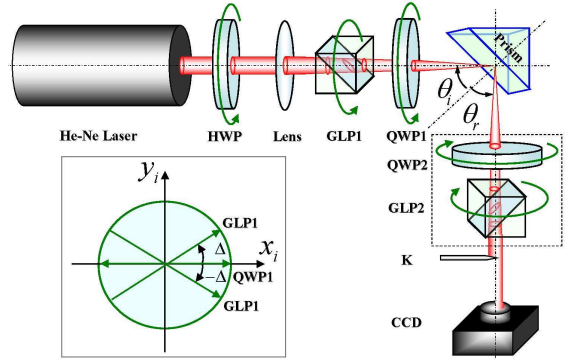}}
\caption{\label{Fig2} (a) Experimental setup for characterizing the
large spin angular splitting on reflection at the Brewster angle.
The light source is a $17\mathrm{mW}$ linearly polarized He-Ne laser
at $632.8\mathrm{nm} $ (Thorlabs HRP170); Prism with refractive
index $n=1.515$ (BK7 at $632.8\mathrm{nm}$ ); Lens, lens with
effective focal length: $50\mathrm{mm}$; HWP, half-wave plate (for
adjusting the intensity); QWP1 and QWP2, quarter-wave plates; GLP1
and GLP2, Glan Laser polarizers; Here, QWP2 together with GLP2 allow
for measuring the Stokes parameter $S_{3}$; K, knife edge (The
purpose of the knife is to produce a single spin accumulation so
that only one spin component can be detected in the CCD); CCD,
charge-coupled device (Coherent LaserCam HR). The inset: The
incident beam is preselected in the left- or right-elliptical
polarization state by GLP1 whose optical axis make angles $\Delta$
or ($-\Delta$) with $x_{i}$-axis. Here, we choose
$\Delta$=$0.5^{\circ}$. }
\end{figure}

We firstly take left-elliptical polarization incident light beam as
an example and the right-elliptical polarization can be obtained in
the similar way. The Jones vector of the left-elliptical
polarization is $(\cos\Delta, i\sin\Delta)^{T}$. Therefore,
according to Eqs.~(\ref{matrixr}) and~(\ref{asi}), we can obtain the
reflected angular spectrum:
\begin{eqnarray}
\tilde{\mathbf{E}}_{r}=\frac{r_p\cos\Delta}{\sqrt{2}}\left[(1+
i\tan\Delta
k_{ry}\delta_{ry}+\eta)\tilde{\mathbf{E}}_{r+}+(1+i\tan\Delta
k_{ry}\delta_{ry}-\eta)\tilde{\mathbf{E}}_{r-}\right].\label{HPT}
\end{eqnarray}
Here, $\delta_{ry}=(1+r_s/r_p)\cot\theta_{i}/k_{0}$ and
$\eta=ik_{ry}\delta_{ry}+r_{s}\tan\Delta/r_{p}$. At any given plane
$z_r=const.$, the transverse displacement of field centroid compared
to the geometrical-optics prediction is given by
\begin{equation}
\delta_{\pm}= \frac{\int\int \tilde{\mathbf{\xi}}^{\ast}_{r\pm}
i\partial_{k_{rx}}\tilde{\mathbf{\xi}}_{r\pm} dk_{rx}
dk_{ry}}{\int\int
\tilde{\mathbf{\xi}}^{\ast}_{r\pm}\tilde{\mathbf{\xi}}_{r\pm}
dk_{rx} dk_{ry}}\label{centroid},
\end{equation}
where $\tilde{\mathbf{\xi}_{r}}_{\pm}$ = $r_p\cos\Delta(1+
i\tan\Delta k_{ry}\delta_{ry}\pm\eta)\tilde{\mathbf{E}}_{r\pm}$. We
note that there needs a theoretical correction and the higher-order
terms should be taken into account when the beam is incident near
the Brewster angle~\cite{Luo2011b}. By making use of a Taylor series
expansion based on the arbitrary angular spectrum component, $r_{p}$
and $r_{s}$ can be expanded as a polynomial of $k_{ix}$:
\begin{eqnarray}
r_{p,s}(k_{ix})=r_{p,s}(k_{ix}=0)+k_{ix}\left[\frac{\partial
r_{p,s}(k_{ix})}{\partial
k_{ix}}\right]_{k_{ix}=0}+\sum_{j=2}^{N}\frac{k_{ix}^N}{j!}\left[\frac{\partial^j
r_{p,s}(k_{ix})}{\partial k_{ix}^j}\right]_{k_{ix}=0}\label{LMD}.
\end{eqnarray}
Using this method, we can obtain the theoretical shift of the single
circular polarization component induced by angular splitting in the
case of left-elliptical polarization:
\begin{equation}
\delta_{\pm}=\pm\frac{2z_{r}r_{s}\frac{\partial
r_{p}}{\partial\theta_i}\left[(k_{0}R+\csc^{2}\theta_{i}-1)\sin2\Delta+\csc^{2}\theta_{i}-1\right]}{k_{0}R\left[2k_{0}Rr_{s}^{2}-(\frac{\partial
r_{p}}{\partial\theta_i})^{2}\right]\cos2\Delta-\left[2k_{0}Rr_{s}^{2}+(\frac{\partial
r_{p}}{\partial\theta_i})^{2}\right](k_{0}R+\csc^{2}\theta_{i}+\cot^{2}\theta_{i}\sin2\Delta-1)}\label{shift}.
\end{equation}
Here $R=k_{0}w_{0}^{2}/2$, $w_{0}$ is the beam waist and $z_{r}$ is
the propagation distance. It should be noted that the reflected
light beam will experience a spatial shift in the case of linear
polarization and a angular displacement according to the elliptical
polarization. In this work, we only consider the elliptical
polarization in which the large spin angular splitting is explored.
\begin{figure}[b]
\centerline{\includegraphics[width=7.5cm]{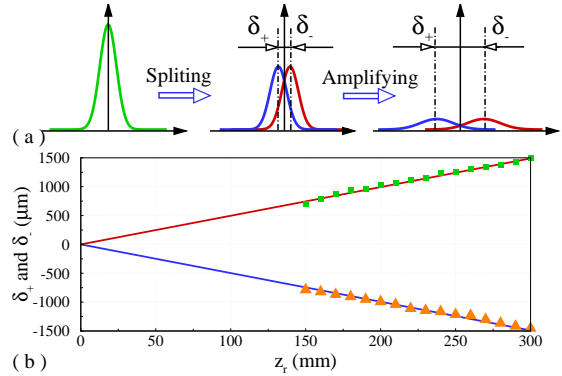}}
\caption{\label{Fig3} (a) Presection and postselection of
polarizations give rise to an amplified spin angular splitting. (b)
Theoretical and experimental results of two spin components induced
by the spin angular splitting for beam waist
$w_0=18.66\mathrm{{\mu}m}$. We measured the $z_{r}$ from the beam
waist. The incident light beam is left-elliptical polarization for
$\Delta$=$0.5^{\circ}$ from the $x_{i}$-axis and incidents at the
Brewster angle $\theta_{i}=56.57^{\circ}$. }
\end{figure}

\section{Weak measurements system}
Next, we focus our attention on the experiment. Figure~\ref{Fig2}
illustrates the experimental setup. A Gaussian beam generated by a
He-Ne laser is preselected as a slightly elliptical polarization
state by GLP1 and QWP1. By choosing the focal length of lens, we can
obtain the desired beam waist in reflection. When the beam impinges
onto the prism interface, the reflected beam is angularly separated
into two opposite spin components. The prism is mounted to a
rotation stage allowing for precise control of the incidence at the
Brewster angle. It should be noted that the reflected mechanism
discussed above can be seen as a built-in post-selection amplified
technique in which the angular splitting is significantly amplified.
To detect the angular splitting, a knife edge is applied to achieve
a distribution of single spin component. We use a CCD to measure the
centroid of the spin accumulation after the knife edge. Using
additional QWP2 and GLP2, we can measure the Stokes parameter
$S_{3}$ which reveals the circular polarization state of the angular
splitting~\cite{Leyder2007}.

The amplifying mechanism of spin angular splitting is schematically
shown in Fig.~\ref{Fig3}(a). The incident beam is preselected in the
elliptical polarization, and then postselected in the circular
polarization state when it reflects on the prism at the Brewster
angle. In our measurement, we first choose the lens with focal
length $f=50\mathrm{mm}$ to generate beam waist
$w_0=18.66\mathrm{{\mu}m}$ and make the incident ligt beam at the
Brewster angle by modulating the GLP1 along to the $x_{i}$-axis.
Then we select the incident light beam as sligltlly left-elliptical
polarization by modulating the $\Delta$=$0.5^{\circ}$ from the
$x_{i}$-axis. Limited by the large holders of the knife edge and
diaphragm, the angular splitting at small propagation distance are
not measured. We measure the displacements every $10\mathrm{mm}$
from $150\mathrm{mm}$ to $300\mathrm{mm}$ [Fig.~\ref{Fig3}(b)]. The
detected splitting value reaches about $1500\mathrm{{\mu}m}$ at the
plane of $z_r=300\mathrm{mm}$. The solid lines represent the
theoretical predictions. The experimental results are in good
agreement with the theory without using parameter fit.

\begin{figure}
\centerline{\includegraphics[width=7cm]{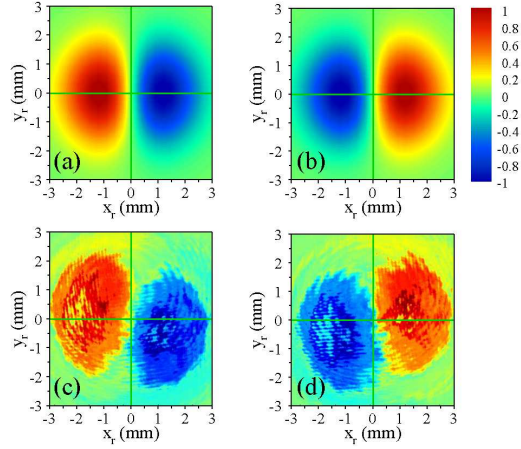}}
\caption{\label{Fig4} Theoretical and experimental results of Stokes
parameter $S_{3}$: (a), (b) Theoretical calculation of Stokes
parameter $S_{3}$ for the left- and right-elliptical polarization;
(c), (d) Experimental results concluded from the intensity
distributions on CCD. The directions of the spin accumulation can be
switched by adjusting the initial handedness of polarizations. Here,
the parameters are the same as that of Fig.~\ref{Fig3}. The
distributions in the plane $z_r=300\mathrm{mm}$ are plotted with
normalized units.}
\end{figure}

To obtain a clear physical picture, it is necessary to analyze the
polarization distribution of the reflected light beam. The Stokes
parameter $S_{3}$ is introduced to describe the circular
polarization state of the spin angular splitting. Here, $S_{3}$=+1
or -1 represents the left- or right-circular polarization.
Figure~\ref{Fig4}(a) and~\ref{Fig4}(b) illustrate the theoretical
polarization distribution of reflected light beam considering the
different left- and right-elliptical polarization incident beam. We
can clearly see that, with the incident beam elliptical polarization
state changing from left (right) to right (left), the spin angular
splitting will reverse the direction. As an analogy of SHE in
electronic system~\cite{Leyder2007}, the directions of spin
accumulations can be switched by the initial handedness of
polarization.

We also carry out another experiment to measure the polarization
distribution described by Stokes parameter $S_{3}$. The experimental
setup is similar to the first experiment. A new Glan laser polarizer
(GLP2) and a new quarter-wave plate (QWP2) are added behind the
prism. The last Glan laser polarizer, quarter-wave plate and CCD
establish a general experimental system for measuring polarization
distribution. By rotating the GLP2 to two angles and holding the
QWP2 along to the $y_r$-axis, we can conclude the Stokes parameter
$S_{3}$ from the intensity distributions on CCD. The rotation angles
are $45^{\circ}$ and $135^{\circ}$, the deviations from the
configuration of the $y_{r}$-axis. The experimental results shown in
Fig.~\ref{Fig4}(c) and~\ref{Fig4}(d) are in good agreement with the
theoretical calculation.

\section{Conclusions}
In conclusion, we have revealed a large spin angular splitting on
reflection at the Brewster angle. The detected splitting reaches
about $1500\mathrm{{\mu}m}$ at $z_{r}=300$$\mathrm{mm}$. As an
analogy of SHE in semiconductor microcavity~\cite{Leyder2007}, we
are able to switch the directions of the spin accumulations by
adjusting the initial handedness of spin states. This phenomenon can
be interpreted from the inversion of horizontal electric field
vector across the Brewster angle. Importantly, we propose a simple
weak measurements system offering an interesting built-in
post-selection technique to explore this angular splitting, which
will provide us a new method on weak measurements technique.

\section*{Acknowledgments} We are sincerely grateful to the anonymous referees,
whose comments have led to a significant improvement of our paper.
This research was partially supported by the National Natural
Science Foundation of China (Grants Nos. 61025024 and 11074068).

\end{document}